\documentclass[aps,prl,reprint,preprintnumbers,nofootinbib]{revtex4-1}
\usepackage{blindtext}

\usepackage{graphicx} 
\usepackage{amsmath}
\usepackage{amsfonts}
\usepackage{amssymb}
\usepackage{physics}
\usepackage{bbold}
\usepackage{natbib}
\usepackage[normalem]{ulem}
\usepackage{orcidlink}

\usepackage[dvipsnames]{xcolor}

\begin{document}
\preprint{MIT-CTP/6088}

\title{Quantum Chinese Remainder Clock}
\date{May 2026}

\author{Ivri Nagar~\orcidlink{0009-0001-2875-3464}}
\email{ivri@mit.edu}
\affiliation{Center for Theoretical Physics -- a Leinweber Institute,
Massachusetts Institute of Technology}
\author{Alioscia Hamma\orcidlink{0000-0003-0662-719X}}
\email{alioscia.hamma@unina.it } 
\affiliation{Scuola Superiore Meridionale, Largo S. Marcellino 10, 80138 Napoli, Italy}
\affiliation{Istituto Nazionale di Fisica Nucleare (INFN), Sezione di Napoli, Italy}
\affiliation{Dipartimento di Fisica `Ettore Pancini', Universit\`a degli Studi di Napoli Federico II, Via Cintia 80126, Napoli, Italy}
\author{Mikel Palmero~\orcidlink{0000-0001-9222-5298}}
\affiliation{Research Laboratory of Electronics, Massachusetts Institute of Technology}
\affiliation{EHU Quantum Center, University of the Basque Country UPV/EHU}
\affiliation{Department of Applied Physics, University of the Basque Country UPV/EHU}

\author{Matthew Radzihovsky}
\affiliation{Department of Electrical Engineering and Computer Science, Massachusetts Institute of Technology}
\author{Shouzhuo Yang ~\orcidlink{0000-0003-0290-5412}}
\email{ysz@mit.edu}
\affiliation{Department of Nuclear Science and Engineering, Massachusetts Institute of Technology}
\author{Seth Lloyd}
\email{slloyd@mit.edu}
\affiliation{Department of 
Mechanical Engineering, Massachusetts Institute of Technology}
\begin{abstract}
       The Chinese remainder theorem is used in metrology for extending the range of quantum clocks/radar/interferometry, where the phase of a signal is known relative to a set of oscillators with different periods. This paper investigates the performance of a quantum-mechanical Chinese remainder clock, consisting of atoms/oscillators with pairwise coprime periods. We provide the optimal initial state and the optimal Heisenberg-limited quantum measurements for measuring time up to the product of the periods. We introduce a novel fault-tolerant post-processing protocol that allows reconstruction of the correct time even in the presence of errors in the remainders. 
\end{abstract}

\maketitle

\noindent\emph{Introduction} – 
``What time is it?"   
Atomic clocks provide an answer to this question with high precision: optical frequency atomic clocks such as NIST's aluminum atom quantum logic clock can measure time to a precision of $10^{-18}$ seconds \cite{Marshall2025}.   For all atomic clocks there is a tension between the precision, given by the inverse frequency of the atoms/oscillators in the clock, and the range: the atoms/oscillators themselves only determine time modulo their period, and we wish to measure time over many periods.   In conventional atomic clocks such as a microwave frequency cesium clock or an optical frequency aluminum clock, the atoms interact with a semiclassical, highly excited oscillator, a maser or laser, whose oscillations we can count: the precision is obtained by using the atoms to monitor the frequency drift of the oscillator.

 The Chinese Remainder Theorem (CRT) arises from a problem posed by Sunzi in his book written between the 3rd and 5th centuries; Aryabhata provided a general algorithm to solve Sunzi's problem in the 6th century.   The theorem states: given $m$ pairwise coprime positive integers $x_j$, the value of a positive integer $x$ between $1$ and $x_1 \ldots x_m$ is uniquely determined by its remainders $x\text{ mod}\,x_j$.
The CRT is useful for extending the range of any system where we can measure the phase within a period of an oscillator, but can't count the periods. The CRT has been profitably applied to atomic clocks, to radar \cite{Xia2007}, and to interferometry \cite{Yankelev2020}.   Heretofore, applications of the Chinese remainder theorem have been classical or semiclassical.  In this paper we propose a fully quantum mechanical CRT quantum clock, in which the underlying quantum systems are harmonic oscillators or ensembles of identical atoms with integer periods $x_j$.     The physics of measuring time using such clocks individually is well-established \cite{Pegg1989, Buzek1999, Holevo1979}.   Here we present the optimal initial states and optimal quantum measurements for applying the CRT to a fully quantum clock.

\smallskip\noindent\emph{Quantum CRT Clock} – Our CRT quantum clock consists of $m$ ``hands'', each a truncated harmonic oscillator with $x_j$ states, $j=1,...,m$, where $x_j$ are chosen pairwise coprime. The Hamiltonian is 
\begin{equation}
H_j = \omega_j \sum_{n=0}^{x_j-1} n |n\rangle_j\langle n| \,, \label{eqn: truncated_single_hamiltonian}
\end{equation}
where $\omega_j = 2\pi/x_j$. Here and throughout, we set $\hbar=1$ and similarly fix our basic time unit to be $1$. Up to degeneracy, this Hamiltonian is the same as for an ensemble of $x_j-1$ identical two-level spins or atoms with frequency $\omega_j$.

Define the phase state $|\varphi\rangle_j$ for the $j$th oscillator 
\begin{align}
    \ket{\varphi}_j = \frac{1}{\sqrt{x_j}} \sum_{n=0}^{x_j-1} e^{-i n \varphi} \ket{n}_j\, . \label{eqn:phase_state_definition}
\end{align}
Over time $t$, $\ket{\varphi}_j$ evolves to $e^{-iH_jt} |\varphi\rangle_j = |\varphi + \omega_j t\rangle_j$, so that a measurement of the phase determines the time mod $x_j$. Phase states in general are not orthonormal, but we can define an orthonormal basis of phase states $\{ |\varphi^j_k\rangle_j \}_{k=0}^{x_j-1}$ for each truncated oscillator, with $\varphi^j_k = 2 \pi k / x_j$ \cite{holevoProbabilisticStatisticalAspects2011}. The corresponding phase operator is then
\begin{equation}
\Phi_j^\text{disc} = \sum_k \varphi_k^j |\varphi^j_k\rangle_j \langle \varphi^j_k|\,. \label{eqn:phase_operator_discrete}
\end{equation}

\begin{figure}[t]
    \centering
    \includegraphics[width=\linewidth]{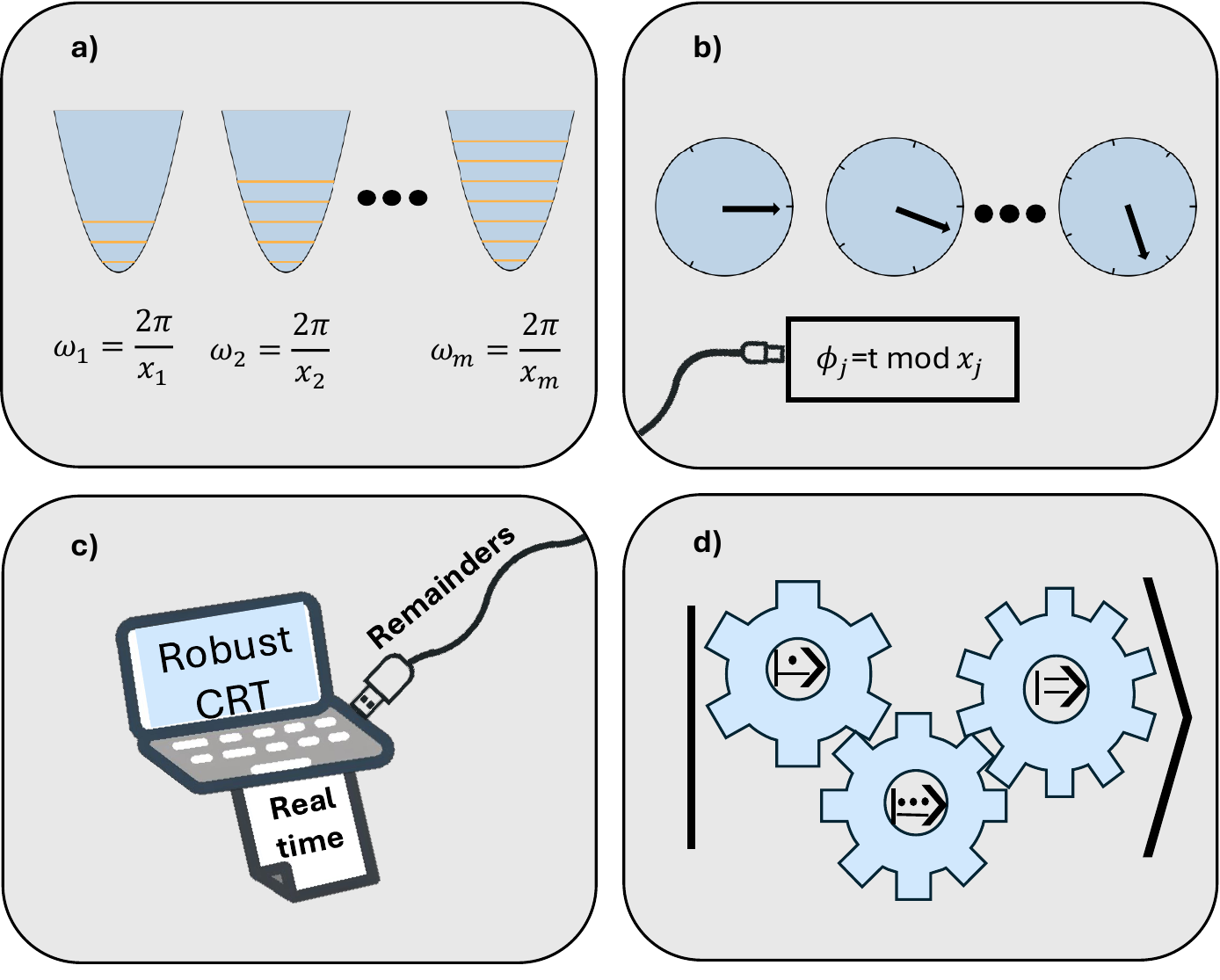}
    \caption{Schematic of the proposed protocol. a) State preparation: We prepare $n$ truncated harmonic oscillators, with coprime periods, in the optimal initial states for phase estimation. b) Measuring the time $t$: We perform the optimal Heisenberg-limited measurement of phase on each oscillator, to find the remainder of $t$ by each period. c) Chinese remainder theorem reconstruction: From the remainders, we recover $t$, using a robust algorithm to overcome potential experimental errors. d) Extended dynamical range: This idea is reminiscent of the structure of the Maya Long Count Calendar. With enough periods (denoted by Maya numerals in kets in the figure), we can extend the dynamical range of our device exponentially. \newline
    \label{fig:schematic}}
\end{figure}

A simple, deterministic protocol then allows us to measure integer times using the Chinese remainder theorem. We initialize each of the oscillators in the $k=0$ state
$\ket{\psi(t=0)}_j=|\varphi^j_{k=0}\rangle_j=|\varphi= 0 \rangle_j$, such that at unit time intervals it ``ticks'' deterministically through the states $\ket{\psi(t=k)}_j=|\varphi^j_k\rangle_j$, $k=1,2,3,  \ldots$, wrapping around at $k=x_j$. 
We now measure $\Phi_j^\text{disc}$ at integer times to determine $\varphi^j_t=2 \pi t / x_j \,(\text{mod } 2 \pi)$, from which we obtain the remainder of $t_j=t \,(\text{mod } x_j)$ for each of the oscillators. 
Since $x_j$ are pairwise coprime, the standard classical algorithm for the CRT then reveals $t$ from these remainders, as long as $t< \prod_{j=1}^m x_j$ (the ``Maya calendar range'' time for the set of hands), see Fig. \ref{fig:schematic}.

This essentially classical protocol, while straightforward, is not particularly satisfactory. In order to make measurements at precisely integer times, our measurement apparatus must possess a frequency-stable classical oscillator to determine when the individual phase measurements are to be made; having such a stable oscillator, we may simply count its periods and use it as a clock, without ever introducing a microscopic quantum clock in the first place. Measuring the phases at non-integer times introduces errors, and a naive reconstruction of $t$ is notoriously sensitive to any error in the remainders,\footnote{
This may initially seem strange, but simple generic examples bear this out. For instance, consider $x_1=5$, $x_2=7$; if we measured $t= 1\,(\text{mod } 7)$ and $t= 3\,(\text{mod } 5)$, then $t=8\,(\text{mod } 35)$, while if instead $t= 1\,(\text{mod } 7)$ but $t= 4\,(\text{mod } 5)$, then $t=29\,(\text{mod } 35)$. Thus changing even a single remainder by $1$ leads to a shift in the result on the order of the entire range.
} necessitating some form of error correction. 
In the following sections we provide a protocol for continuous time measurement that employs optimal quantum measurements for each oscillator and provides both the Heisenberg quantum limit and extended dynamical range.

%
%

\smallskip\noindent\emph{Continuous time and measurement} – The optimal initial states and measurement procedures for measuring
time continuously using truncated harmonic oscillators or ensembles of atoms are known \cite{Holevo1979, Pegg1989, Buzek1999}. The continuous phase operator for each oscillator is 
\begin{equation}\label{eqn:phase_operator_cont}
\Phi_j^{\mathrm{cont}} = \int_{-\pi}^{\pi} \frac{d\varphi}{\omega_j}~ \varphi |\varphi\rangle_j\langle \varphi| \,,
\end{equation}
where the normalization of the measure is necessary to ensure this is a valid POVM, $\int_{-\pi}^{\pi} \frac{d\varphi}{\omega_j}~ |\varphi\rangle_j\langle \varphi|=\mathbb{1}_j$.  Holevo \cite{Holevo1979} showed that this operator gives the optimal continuous phase measurement
, yielding the minimum variance in phase for any initial state.
The optimal initial state is given by Buzek \textit{et al}  \cite{Buzek1999},
\begin{align}
    \left|\psi_{\text{opt}}\right\rangle_j = \frac{\sqrt{2}}{\sqrt{x_j}} \sum_{n=0}^{x_j-1} \sin \frac{\pi(n+1 / 2)}{x_j}|n\rangle.
\end{align}

After evolving from $\left|\psi(t=0)\right\rangle_j =\left|\psi_{\text{opt}}\right\rangle_j $ for time $t$, the probability density of measuring $\varphi_j$ is
\begin{align}
\label{eqn:continuous_optimal_measure_distribution}
    P_{\text{opt}}(\varphi_j\mid t) =&
    \frac{2 \sin^2\Bigl(\tfrac{\pi}{2x_j}\Bigr) \cos^2\!\Bigl(\tfrac{x_j\delta_j}{2}\Bigr)\,
      \cos^2\!\Bigl(\tfrac{\delta_j}{2}\Bigr)}
     {x^2_j \sin^2\!\Bigl(\tfrac{\delta_j+\pi/x_j}{2}\Bigr)\,
      \sin^2\!\Bigl(\tfrac{\delta_j-\pi/x_j}{2}\Bigr)}\,,\\
      & \text{where} \qquad \delta_j=\varphi_j-\omega_j t\,.\nonumber
\end{align}
In particular, this distribution has a sharp central peak at $\bar \varphi_j =\omega_j t~\mathrm{mod}~2\pi$. The width of this peak,  which is the standard deviation of $\varphi_j$ when restricted to lie between $\bar \varphi_j-\pi$ and $\bar \varphi_j+\pi$, is $\Delta \varphi_{j,\,\text{opt}} \simeq \frac{\pi}{x_j}$ at large $x_j$.
Thus each phase measurement with this optimal initial state yields a measurement $\tilde t_j=\varphi_j/\omega_j \,(\text{mod } x_j)$ of the time remainder $t_j$ for the $j$'th oscillator, with Heisenberg scaling error $\Delta t_j \simeq 1/2$. 

We now provide a scheme to deal with measurement errors.\footnote{
The error correction protocols put forth in this work are classical, in that we first measure the phases and then perform corrections. In principle, one could also consider more fundamentally quantum error correction, wherein quantum processing is done to the superposition of oscillator states themselves and the measurement is postponed to the end. Such quantum protocols may be able to achieve results strictly better than classically possible \cite{ChaTil23, CheLiu22}; we leave such interesting extensions to future work.
} Our prescription for the fractional part of the remainders is to round them, use the CRT for the integer part, and add back the fractional part at the end. As described earlier, the traditional CRT reconstruction is generically volatile, and accordingly we are interested mostly in ensuring that the determination of the integer part of the time is correct, as the fractional part is less sensitive and can be recovered in a straightforward manner. Our method affords an exact measurement of the integer part of the time with exceedingly high probability, at the cost of adding states to the Hilbert space, while incidentally also improving the resolution of the fractional part. 
We note that there are various other robust reconstruction algorithms known for the CRT, see \cite{Xia2007, Li2009, Xu2018, XIAO2015242, xiaoErrorCorrectionPolynomial2015, Shpar2004} and references therein; our proposal is particularly suited to the problem at hand: it explicitly deals with the fractional parts, is robust against a physical error distribution, requires only a modest increase of the oscillator Hilbert space dimension, and is simple and computationally inexpensive. 

\smallskip\noindent\emph{Increasing phase resolution} – Each truncated harmonic oscillator measures time to an accuracy of order $\pm 1/2$.   We can improve the accuracy of the measurement of time by increasing the truncation scale:  we keep the same
frequency $\omega_j$ for the $j$'th oscillator, but increase the number of states to $Zx_j$, where $Z$ is an integer.  We use the Buzek optimal initial state for each oscillator.  Each oscillator now ``ticks'' $Z$ times more frequently, and the phase measured for each oscillator has uncertainty that scales inversely with $Z$, namely that $\Delta \varphi_j = \pi/Zx_j$ and
$\Delta t_j = 1/2Z \equiv \Delta t$ for each oscillator.
From equation (\ref{eqn:continuous_optimal_measure_distribution})
, for large $Z$ the probability that the measured $\varphi_j$ is more than $\epsilon$ away from $\omega_j t$ is
\begin{align}\label{eqn:failure}
P_\text{opt}\!\left(|\delta_j|\ge \epsilon\right)
\simeq
\frac{4\pi}{3\,(Zx_j\epsilon)^3} \,.
\end{align}
We thus can attain any desired resolution with any desired probability by increasing $Z$ sufficiently.

\bigskip\noindent{\em Protocol} -- 

\smallskip\noindent(1) Begin
with the optimal initial state for each oscillator, evolve for an unknown time $t< x_1 \ldots x_m$, and measure the phases $\phi_j$ using the optimal Holevo covariant measurement to yield estimates $\tilde t_j$ of time $t_j=t \,(\text{mod } x_j)$ for each oscillator. The standard deviation in the value of each $\tilde t_j$ is of order $1/2Z$, where $Z$ is the truncation scale multiplier, which we shall choose to be sufficiently large as explained below.

\smallskip\noindent(2)
The distribution of measured times $\tilde t_j$ follows Eq. \eqref{eqn:continuous_optimal_measure_distribution}.  
We begin by considering all non-integer parts $r_j = \tilde t_j - \lfloor \tilde t_j \rfloor$ of the values $\tilde t_j$. Suppose now that $|t_j-\tilde t_j|<\frac{1}{4}$ for all $j$ (taken $\text{mod } x_j$), as will happen with high probability for large $Z$, so that all measured $\tilde t_j$ lie in an interval of length at most $1/2$.

If $\max_j{r_j}-\min_j{r_j}<1/2$, then we may simply round each measured $\tilde t_j$ down, to $t'_j = \lfloor \tilde t_j \rfloor$. However, if $\max_j{r_j}-\min_j{r_j}>1/2$, then evidently the measurements ``wrap around'' an integer boundary, and so we must round fractional values below $1/2$ down and above $1/2$ up. That is, in this case we round each $\tilde t_j$ to the nearest integer $t'_j = \lfloor \tilde t_j \rceil$. In either case we then identify the $r'_j=\tilde t_j-t'_j$ as the non-integer correction, which we estimate simply to be $r'=(1/m) \sum_j r'_j$.

\smallskip\noindent(3)
Calculate the integer value $t'$ of $t$ by applying the Chinese remainder theorem inversion to the $t'_j$.  Our final estimate of $t$ is then $t' +  r'$.   

\smallskip\noindent
We now return to the matter of choosing $Z$. The integer part of the time is correctly reconstructed as long as the assumption in step (2) holds: the probability of failure is bounded by the probability that $|t_j-\tilde t_j  |\ge 1/4$ for some $j$. This may be computed directly from \eqref{eqn:continuous_optimal_measure_distribution}; at large $Z$, from Eq. \eqref{eqn:failure}, we have that $P_\text{opt}\!\left(|t_j-\tilde t_j  |\ge 1/4\right)
\simeq
\frac{32}{3\pi^2Z^3}$, and so for the method to succeed with probability $p=1-\epsilon$ we choose 
\begin{equation}\label{eqn: Z bound}
    Z\geq Z_{\text{min}} \simeq \left[\frac{32}{3\pi^2}\frac{1}{1-p^{1/m}} \right]^{\frac1 3}\simeq\left(\frac{m}{\epsilon} \right)^{\frac1 3}.
\end{equation}
Then, the uncertainty on the final value follows solely from that on the measured fractional parts, in the standard way. The unbiased estimator of their standard deviation is $ \Delta^2 = \sum_j (r'_j-r')^2/(m-1)$, yielding an overall uncertainty $\Delta/\sqrt m \approx 1/(2Z \sqrt m)$ for the estimate of $t$.  

\begin{figure}[t]
\includegraphics[width=\columnwidth]{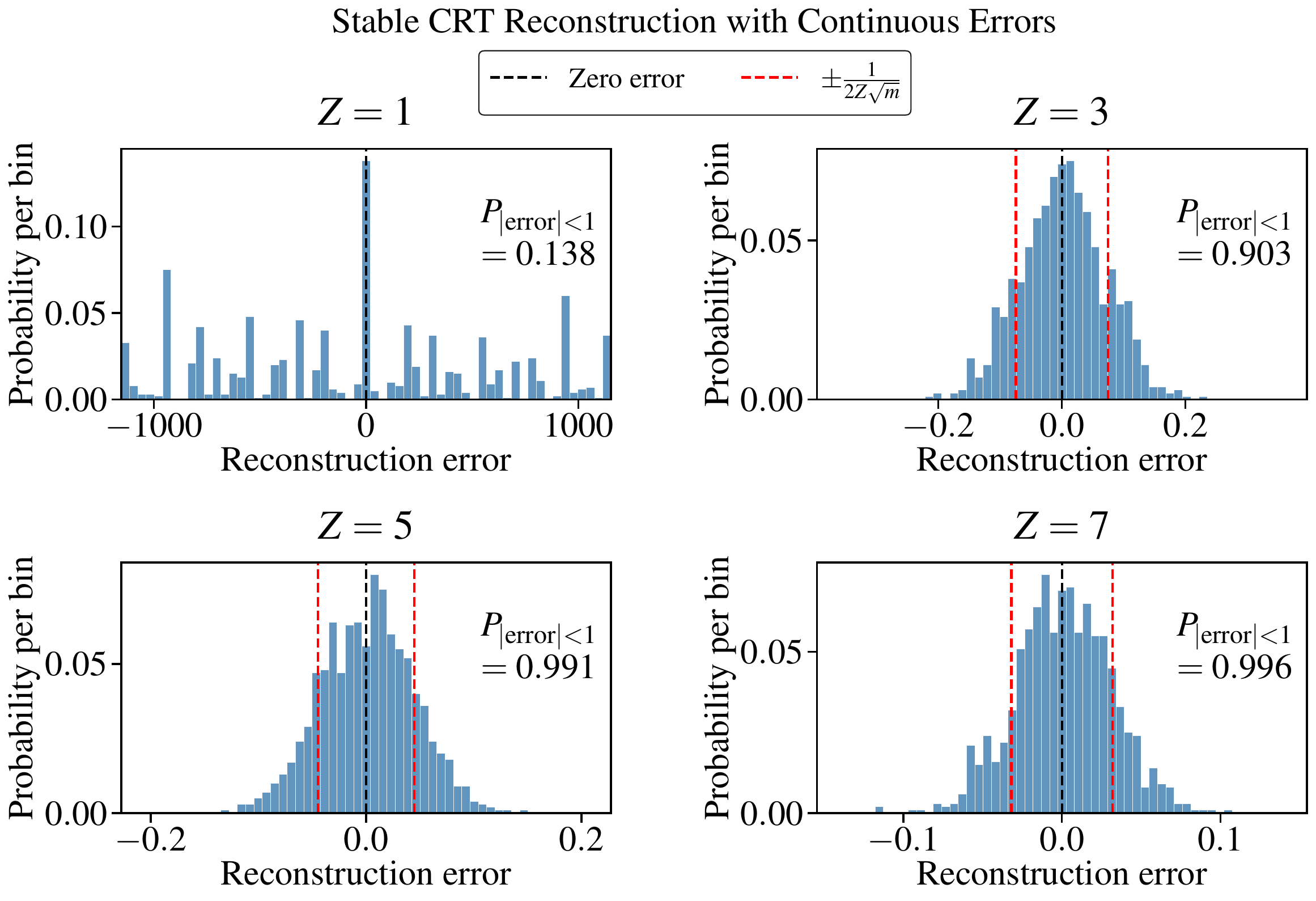}
\caption{\label{fig:Traditional_CRT_Simulation} Reconstruction errors for the proposed protocol with $\{x_j\} = \{2, 3, 5, 7, 11\}$ and $Z \in \{1,3 ,5, 7\}$, for $N=1000$ trial times chosen uniformly at random from $[0, 2310)$ \cite{yserene_2026_21827736}. 
For $Z=1$ (no error correction), the error is distributed across the entire range, while for larger $Z$ the distribution is peaked around $0$ with uncertainty $\frac{1}{2Z\sqrt{m}}$ (red dashed lines) -- 
e.g. $Z=5$ is sufficient to ensure error less than $1$ in $~ 99\%$ of cases.
}
\end{figure}

Figure \ref{fig:Traditional_CRT_Simulation} illustrates a simulation of the above protocol, demonstrating significant improvement in the reconstruction results as we increase $Z$. As discussed, the most critical failure mode is not a small continuous error, but an incorrectly rounded remainder, which can produce a reconstruction error of order $\prod_j x_j$. This behavior is visible at small $Z$, where the reconstruction exhibits broad catastrophic failures, spread sporadically through the entire range. As $Z$ increases, this failure mode rapidly becomes increasingly rare, and the remaining error is due to the fractional part, with narrow distribution around zero. The simulation therefore confirms the main role of increasing $Z$. 

%
%
%
%

\smallskip\noindent\emph{Proof of optimality} -- The procedure we have given here for implementing a fully quantum CRT clock employs states and measurements that are not entangled between the different hands. For each hand, the initial state $|\psi(0)\rangle$ is indeed highly non-classical: it is anti-squeezed in energy for each truncated oscillator, or equivalently an entangled spin anti-squeezed state for each atomic ensemble. It might be suspected that a more accurate procedure for measuring time, using states and measurements that are entangled between the hands, is possible.

Surprisingly, this is not the case, as we now prove: entanglement does not help.
Indeed, the optimal measurement of time for any Hamiltonian is given by the Holevo covariant measurement \cite{Holevo1979} -- the 
POVM elements are projections onto the states $|\chi(t)\rangle = e^{-iHt} |\chi(0)\rangle$, where $|\chi(0)\rangle$ is a uniform superposition over energy eigenstates of $H$. In our case, these states are simply the unentangled states
$|\phi(t)\rangle_1 \otimes \ldots \otimes |\phi(t)\rangle_m$, the tensor product of the phase states for the individual oscillators.

The Holevo procedure does not in general specify the optimal initial state;   
consider, therefore, an arbitrary initial state $|\chi(0)\rangle_{1\ldots m}$ (entangled between the hands). Let the reduced density matrix for the $j$'th oscillator be $\chi_j(0) = \sum_k
p_k^j |\mu_k(0)\rangle_j\langle\mu(0)_k|$.
This state evolves to
$
\chi_j(t) 
= \sum_k
p_k^j |\mu_k(t)\rangle_j\langle\mu(t)_k|,$
where $|\mu_k(t)\rangle = e^{-iHt} |\mu_k(0)\rangle$. Now, the variance $\Delta_k^2$
of the phase measurement on each individual
$|\mu_k(t)\rangle$ is greater than or equal to the optimal variance $\Delta_{\text{opt}}^2$, obtained when the initial state is the optimal Buzek state for that oscillator.  The variance of any observable $O$ in the state $\chi_j(t)$ obeys $\Delta_\chi^2 = \sum_k p_k \Delta^2_k + \sum_k p_k ( \tr (O \mu_k) - \tr(O\chi))^2$, so since $\Delta^2_k \ge \Delta_{\text{opt}}^2$, we conclude $\Delta_\chi^2\ge \Delta_{\text{opt}}^2$. Hence, the variance in time measurement for any initial state is at least as great as that for the unentangled procedure given above. 

\smallskip\noindent{\it Experimental Realizations and Platforms} --
The quantum CRT clock can be implemented across a range of existing quantum sensing platforms that already confront the fundamental tradeoff between phase sensitivity and dynamic range. In cold-atom interferometry, phase measurements are intrinsically $2\pi$-periodic, limiting the unambiguous range of inertial and gravitational sensing. Recent experiments have addressed this limitation by combining interferometers with different interrogation times or scale factors, producing moiré-like phases with an extended effective period and enabling large enhancements in dynamic range \cite{Yankelev2020}. More generally, composite and multi-interrogation-time protocols combine modular phase measurements to reconstruct signals over a wider range. Related strategies based on large-momentum-transfer atom optics and multi-frequency control further extend the accessible phase accumulation and coherence time \cite{Chiow2011,Kessler2014}. These approaches demonstrate that multi-scale phase estimation is experimentally viable, but they remain fundamentally semiclassical, relying on separate measurements and classical post-processing.

In parallel, rapid progress in quantum metrology has enabled the use of entangled states and optimized measurements to approach fundamental limits of precision. Programmable quantum sensors based on trapped ions have demonstrated near-optimal phase estimation over finite dynamic ranges by jointly optimizing input states and measurement protocols \cite{Marciniak2022}, while entanglement-enhanced interferometry has achieved sensitivities beyond the standard quantum limit \cite{Hosten2016, Cao2024, Zaporski2025}. Compact and deployable quantum sensors are also emerging as practical technologies for real-world applications \cite{Bongs2019}. The present work provides a unifying framework that combines these directions: it extends multi-scale interferometric ideas into the fully quantum regime, where multiple ``moduli'' are treated as a single coherent system and measured optimally. This suggests a new class of quantum sensors that simultaneously achieve large dynamic range and Heisenberg-limited precision, bridging the gap between classical range-extension techniques and quantum-enhanced metrology.

\bigskip\noindent{\it Acknowledgements:} This work was done as a class project for MIT course 2.S372, Design of Quantum Algorithms.   We thank Aparna Gupte, Pamela Pajarillo, Alexander Ungar and Alexander Schmidhuber for helpful suggestions.   SL was supported in part by the U. S. Army Research Laboratory and the U. S. Army Research Office under contract/grant number W911NF2310255, and by DoE under contract, DE-SC0012704.  S. Y. is supported by the Manson Benedict (1932) Fellowship and the Theos J. Thompson Memorial Fellowship from the Department of Nuclear Science and Engineering at MIT. M. P. is supported by the Spanish Ministry of Science, Innovation and Universities through the Jos\'e Castillejo mobility grant (CAS24/00344), and projects PID-2021-126277NB-I00, PID2024-156808NB-I00, the Basque Government through project IT1470-22, the UPV/EHU through project EHU-N25/38, and the EU Flagship on Quantum Technologies through OpenSuperQ+100 (Grant No. 101113946). IN is supported by the Department of Physics at MIT.

\bibliographystyle{apsrev4-1} 
\bibliography{bibliography}

\end{document}